\documentclass[conference]{IEEEtran}
\IEEEoverridecommandlockouts

\usepackage{cite}
\usepackage{amsmath,amssymb,amsfonts}
\usepackage{graphicx}
\usepackage{textcomp}   
\usepackage{xcolor}
\usepackage{soul}
\definecolor{myorange}{RGB}{230,120,20}

\begin{document}

\title{Design and Development of an ML/DL Attack Resistance of RC-Based PUF for IoT Security}

\author{%
\IEEEauthorblockN{Joy Acharya, Smit Patel, Paawan Sharma, Mohendra Roy*}
\IEEEauthorblockA{Department of ICT, Pandit Deendayal Energy University\\
Gandhinagar, Gujarat 382007, India\\
\texttt{*Corresponding author: mohendra.roy@ieee.org}}
\thanks{Corresponding author: Mohendra Roy.}
}

\maketitle

\begin{abstract}
Physically Unclonable Functions (PUFs) serve as promising hardware security tools for Internet of Things (IoT) applications, where device authentication must be both fast and reliable. The randomness and uniqueness characteristics of PUF make them suitable for the authentication of IoT in resource-constrained environments. However, the recent progress in Machine Learning (ML) and Deep Learning (DL) techniques has raised some serious concerns about PUF security, as modeling attacks can learn and copy challenge--response patterns with high accuracy, which puts the security of PUFs at risk. In this work, we present a custom-developed resistor-capacitor (RC) based dynamically reconfigurable PUF operating with 32-bit challenge \& response pairs (CRPs), that can resist ML/DL attacks. To support this, we have characterized the resistance of our RC-PUF against a systematic adversarial attack. We generated a dataset of CRPs of 32 bits from the PUF and organized it into training, validation, and test sets. To test whether the CRPs of the PUF could be modeled, we applied several well-known machine learning techniques, including Artificial Neural Networks (ANN), Gradient Boosted Neural Networks (GBNN), Decision Trees (DT), Random Forests (RF), and Extreme Gradient Boosting (XGBoost), to test their ability to replicate PUF behavior. At the time of training all these models have achieve 100\% accuracy. However, their performance on unseen test data was close to random guessing, with accuracies of 51.05\% (ANN), 53.27\% (GBNN), 50.06\% (DT), 52.08\% (RF), and 50.97\% (XGBoost). These shows that the ML models are failed to learn the CRPs from the proposed PUF. These results show the strong resistance of the proposed dynamically reconfigurable PUF to ML-driven modeling attacks, as even advanced methods fail to reproduce accurate response patterns. This work gives clear evidence that proposed PUF architectures improve robustness and resistance to adversarial attack (such as ML attacks). Therefore, the selected PUF systems are effective for securing next-generation IoT authentication against machine learning--based threats with minimal resource overhead. This simple RC-PUF can be an alternative to the costly and resource hungry encryption methods for IoT Security.
\end{abstract}


\section{Introduction}
IoT connects billions of devices, many of which cannot support heavy cryptographic methods \cite{al2023physical}. PUFs provide a lightweight alternative, as they exploit unavoidable manufacturing variations in integrated circuits to generate unique responses for authentication \cite{panchal2022mini}. These microscopic differences are beyond the control of the manufacturer, giving each device a distinct fingerprint \cite{al2023physical}. This removes the need to store secret keys in memory and makes PUFs well suited for IoT security. Our custom-developed resistor-capacitor (RC)-based PUF is particularly advantageous for IoT \cite{acharya2025optimized}, as it relies on simple passive RC components that consume ultra-low power, occupy minimal area, and scale easily while maintaining high analog entropy, making it compact, energy-efficient, and highly secure for resource-constrained edge devices. Unlike conventional Ring-Oscillator, SRAM, or Arbiter PUFs that depend on digital timing races or memory startup states, our RC design derives uniqueness from intrinsic analog variability in RC time constants caused by fabrication mismatches. This analog entropy ensures decorrelated, non-learnable challenge--response mappings, directly contributing to the enhanced attack resistance of the proposed RC-PUF. In this work, we examine the resistance of a reconfigurable RC-PUF against ML and DL attacks. We generated 32-bit CRPs and tested models including ANN, GBNN, DT, RF, and XGBoost \cite{kumar2018machine}. While these models fit the training data almost perfectly, their accuracy on unseen data was close to random guessing, confirming the strong robustness of our RC-PUF design for IoT authentication. A PUF can be regarded as having strong resistance to machine-learning attacks if a trained ML model, even with 100\% training accuracy, achieves a test accuracy close to 50\%. Such performance is equivalent to random guessing, indicating that the model cannot reliably predict the corresponding PUF response bits (of 32 bits).

\section{Methodology}
This work aimed to assess the IoT-oriented security of a custom reconfigurable RC-based PUF capable of resisting ML and DL modeling attacks. The experimental study followed a structured pipeline encompassing hardware characterization, dataset generation, and model-based evaluation.

\subsection{Dataset Generation}
The dataset was generated from our custom-developed, hardware-implemented RC-PUF prototype, as detailed in \cite{acharya2025optimized}. Using randomly generated 32-bit challenges, the system produced corresponding 32-bit responses under multiple operating configurations, including First-Order and Second-Order RC-PUF architectures, with and without UID integration, and pulse widths of 2~\textmu s and 32~\textmu s. For each configuration, 80,000 challenge--response pairs (CRPs) were collected, ensuring sufficient diversity and statistical reliability for subsequent machine learning and deep learning--based analysis.

\begin{figure*}[t]
\centering
\includegraphics[width=0.65\linewidth, height=5.5cm]{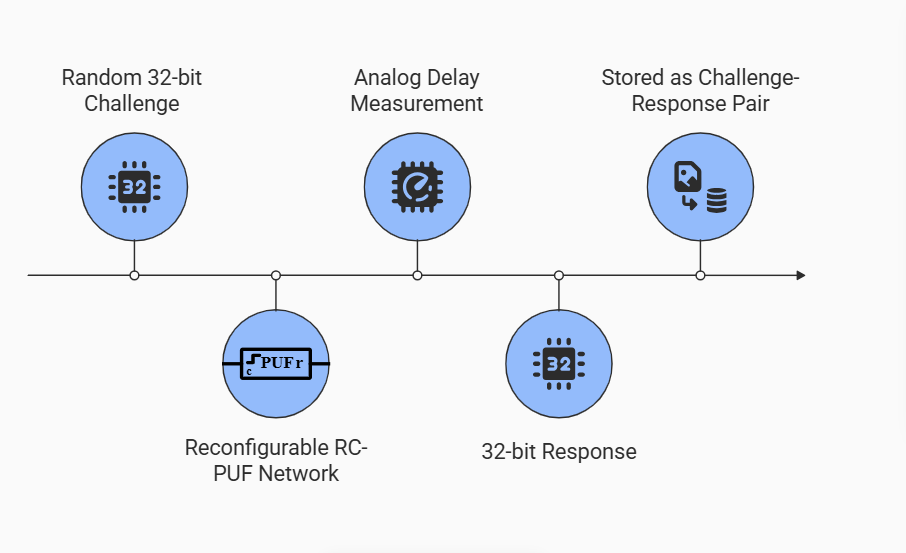}
\caption{RC-PUF dataset generation workflow showing 32-bit challenge 
input, analog delay measurement, and digital response storage.}
\label{fig:rcpuf-architecture}
\end{figure*}

As illustrated in Fig.~\ref{fig:rcpuf-architecture}, each dataset sample is generated through a fixed and repeatable hardware evaluation flow. A 32-bit digital challenge is first applied to the reconfigurable RC-PUF network, where it determines the active RC paths and signal propagation characteristics. The resulting analog delay, influenced by intrinsic process variations and configuration-dependent behavior, is measured using on-chip timing circuitry. This analog measurement is then digitized to generate a stable 32-bit response corresponding to the applied challenge.

The resulting CRP's are stored sequentially without post-processing or feature engineering, preserving a true black-box observation model. By maintaining identical acquisition conditions across all configurations and operating modes, the dataset ensures consistency while capturing meaningful variations induced by RC order, pulse width, and UID presence. This structured yet hardware-faithful dataset generation process provides a reliable foundation for evaluating the robustness of RC-PUF architectures against data-driven modeling attacks. We have divided the dataset into 70:20:10 ratios for training, validation, and testing. The evaluation of the PUF is conducted to assess its uniqueness, reliability, uniformity, bit-aliasing, and randomness, as described in our previously published work \cite{acharya2025optimized}.

\subsection{Machine Learning Models}
Several classical ML models were used to evaluate how vulnerable the PUF is to modeling attacks based on the CRPs dataset. Each challenge vector is represented as $x \in \{0,1\}^{32}$ and its corresponding response as $y \in \{0,1\}^{32}$. Each bit of $y$ is treated as an individual subtask, and the models were trained to predict 32 outputs in parallel, using binary cross-entropy loss across tasks. A PUF is an ML attack resist if the CRP cannot be learned by any model, and the prediction accuracy is not beyond 50\%. In this study, we trained the ML models for its hightest accuracy (100\%). This is to ensure that the model learn the CRP well. Then we test the trained model on the test dataset (from the same PUF).

\textbf{Decision Tree (DT):}
As a baseline, DTs were used where each of the 32 CRP's bits was trained with a separate tree, giving a model set $M=\{DT_1,\ldots,DT_{32}\}$. During inference, the same 32-bit challenge is given to all trees; each $DT_i$ produces its prediction for bit $y_i$, and finally all predictions are concatenated to form the complete 32-bit response vector $\hat{y}$. Tree depths from 1 to 20 were tested. Training accuracy reached 100\% at depth 15, while validation accuracy stayed around 50.37\% with 0\% exact matches \cite{huang2024pfo,vanneschi2023decision}. This shows DTs memorize well but generalize poorly.

\textbf{Random Forest (RF):}
RF extend DTs by combining multiple randomized trees under bagging. In our setup, one RF model was trained per response bit, giving $M=\{RF_1,\ldots,RF_{32}\}$. For a given challenge, each RF uses majority voting across its trees to decide the bit prediction. The outputs of all 32 RFs are then combined into the predicted response. We varied the number of trees from 1 to 35; training accuracy reached 100\% at 33 trees, but validation and test bitwise accuracies were only about 51.5\%, with exact-match accuracy at 0\% \cite{bhatta2024advancing}.

\textbf{Extreme Gradient Boosting (XGBoost):}
XGBoost improves on DTs by training shallow trees in sequence, where each new tree corrects the residual errors of the previous ones. Again, we trained one XGBoost model per response bit, $M=\{XGB_1,\ldots,XGB_{32}\}$. At inference, all 32 models are applied to the challenge in parallel, and their outputs are concatenated into a response. We tested boosting rounds from 1 to 35; training accuracy reached 100\% at 35 rounds, while validation and test accuracies remained around 51.3\%, with 0\% exact-match accuracy \cite{9526634}.

\subsection{Deep Learning Models}
To complement the classical ML approaches, deep learning models were applied to learn the nonlinear CRPs mapping of the PUF. Both models were trained on dataset, where each challenge vector $x \in \{0,1\}^{32}$ was mapped to its 32-bit response $y \in \{0,1\}^{32}$.

\textbf{Artificial Neural Network (ANN).}
We constructed an ANN to map 32-bit challenges to 32-bit responses. The architecture used hidden layers of sizes [5000, 2048, 1024, 512, 256, 64] and an output layer of 32 units. LeakyReLU activations were applied, with batch normalization and dropout (0.2) in the first three layers for stability \cite{9526634}. The model was trained with Adam ($\text{lr}=10^{-3}$, weight decay $10^{-4}$) and a learning-rate scheduler. BCEWithLogitsLoss \cite{li2024rediscovering} was used, with predictions obtained via a 0.5 Sigmoid threshold. Metrics included bitwise accuracy, Hamming loss, and exact-match accuracy \cite{shi2019approximation}. The ANN quickly reached 100\% training accuracy but only \textasciitilde51\% on the test set, with 0\% exact matches.

\textbf{Gradient Boosting Neural Network (GBNN).}
We also experimented with a GBNN \cite{emami2023sequential}, where multiple ANNs are trained sequentially, each correcting the residuals of the previous ones. After $T$ stages, the combined model is $F_T(x) = F_0(x) + \sum_{t=1}^T \rho_t h_t(x)$. Each base learner was an 8-layer ANN, trained with BCEWithLogitsLoss and the Adam optimizer ($\eta=0.01$, weight decay $10^{-5}$). By stage 10, the ensemble reached 100\% training accuracy and 53.27\% test accuracy, but still 0\% exact matches. Bitwise accuracy improved slightly over a single ANN, yet the boosted networks failed to generalize, leaving strong RC-PUFs resistant.

\subsection{Error Functions}
In binary classification, two commonly used error functions are those based on the binary cross-entropy (BCE). The deep models were trained with the more numerically stable in this work.

\begin{table}[htbp]
\caption{Error functions used in training.}
\label{tab:error_functions}
\centering
\renewcommand{\arraystretch}{1.3}
\setlength{\tabcolsep}{6pt}
\begin{tabular}{|p{0.30\linewidth}|p{0.60\linewidth}|}
\hline
\textbf{Error Function} & \textbf{Equation} \\
\hline
Binary Cross-Entropy (BCE) &
$\mathcal{L}_{\text{BCE}} = - \frac{1}{32N} \sum_{i=1}^N \sum_{j=1}^{32} \big[ y_{ij} \log \hat{p}_{ij} + (1-y_{ij}) \log (1-\hat{p}_{ij}) \big]$ \\
\hline
BCEWithLogitsLoss &
$\mathcal{L}_{\text{BCE+Logits}} = - \frac{1}{32N} \sum_{i=1}^N \sum_{j=1}^{32} \big[ y_{ij} \log \sigma(\hat{z}_{ij}) + (1-y_{ij}) \log (1-\sigma(\hat{z}_{ij})) \big]$ \\
\hline
\end{tabular}
\end{table}

BCE requires us to apply the raw outputs (logits) $\hat{z}_{ij}$ first to a sigmoid function to convert them into a probability $\hat{p}_{ij}$. The loss can be computed only at that time. Conversely, the loss computation and sigmoid are calculated simultaneously on BCE With Logits Loss. It operates on the raw logits $\hat{z}_{ij}$, directly, avoiding additional computation as well as making it more stable. It is due to this that the loss function we used in our deep learning experiments is it.

\subsection{Evaluation Metrics}
To assess model performance, three standard metrics were used bitwise accuracy, Hamming loss, and exact match accuracy.

\begin{table}[htbp]
\caption{Evaluation metrics for PUF modeling.}
\label{tab:evaluation_metrics}
\centering
\renewcommand{\arraystretch}{1.3}
\setlength{\tabcolsep}{6pt}
\begin{tabular}{|p{0.30\linewidth}|p{0.60\linewidth}|}
\hline
\textbf{Metric} & \textbf{Equation} \\
\hline
Bitwise Accuracy &
$\text{Acc}_{\text{bit}} = \tfrac{1}{32N} \sum_{i=1}^{N} \sum_{j=1}^{32} \mathbb{I}[\hat{y}_{ij} = y_{ij}]$ \\
\hline
Hamming Loss &
$\text{HL} = \tfrac{1}{32N} \sum_{i=1}^{N} \sum_{j=1}^{32} \mathbb{I}[\hat{y}_{ij} \neq y_{ij}]$ \\
\hline
Exact Match Accuracy &
$\text{EM} = \tfrac{1}{N} \sum_{i=1}^{N} \mathbb{I}[\hat{\mathbf{y}}_i = \mathbf{y}_i]$ \\
\hline
\end{tabular}
\end{table}

Bitwise accuracy is a fraction of the bits of a response that are correctly predicted even when using all the CRPs and Hamming loss is the fraction of bits incorrectly predicted. The strictest metric is exact match accuracy which involves having all 32 bits of response in a CRP correct at the same time. Collectively, these measures deliver an even-handed perception of per-bit prediction quality and fidelity of response broadly.

\section{Results and Discussion}
The resistance of the proposed RC-PUF against data-driven modeling attacks was evaluated using five representative machine learning and deep learning models: Artificial Neural Network (ANN), Gradient Boosted Neural Network (GBNN), Decision Tree (DT), Random Forest (RF), and XGBoost. For each model, training, validation, and test performance were analyzed to assess convergence behavior, generalization capability, and susceptibility to modeling attacks. All models were trained using identical datasets and evaluation protocols to ensure a fair comparison.

Across all evaluated models, near-perfect training accuracy ($\approx 100\%$) was observed, confirming that the models possess sufficient capacity to memorize the training challenge--response pairs (CRPs). However, validation accuracy consistently remained close to the random-guess baseline, ranging between 50\% and 53\%. This sharp discrepancy between training and validation performance indicates severe overfitting and demonstrates that the learned representations fail to generalize to unseen CRPs. Such behavior suggests the absence of an exploitable functional relationship between challenges and responses.

A quantitative summary of the modeling attack performance is provided in Table~\ref{tab:perf_only}. None of the evaluated models achieved meaningful validation accuracy, and the exact-match rate remained zero across all cases. Although GBNN attained the highest validation accuracy (53.27\%), the marginal improvement over random guessing is statistically insignificant and does not translate into successful response prediction. These results confirm that the RC-PUF mapping remains effectively non-learnable under conventional ML/DL attack models.

\begin{table}[htbp]
\centering
\caption{Model performance summary.}
\label{tab:perf_only}
\renewcommand{\arraystretch}{1.2}
\setlength{\tabcolsep}{6pt}
\begin{tabular}{|l|c|c|c|}
\hline
\textbf{Model} 
& \textbf{\shortstack{Train Accuracy\\(\%)}} 
& \textbf{\shortstack{Val Accuracy\\(\%)}} 
& \textbf{\shortstack{Exact Match\\(\%)}} \\
\hline
ANN      & 100 & 51.06 & 0.00 \\
GBNN     & 100 & 53.27 & 0.00 \\
XGBoost  & 100 & 50.97 & 0.00 \\
DT       & 100 & 50.37 & 0.00 \\
RF       & 100 & 51.32 & 0.00 \\
\hline
\end{tabular}
\end{table}

To further analyze the statistical behavior of predicted responses, bit-level metrics including mean, variance, and entropy were computed and are summarized in Table~\ref{tab:bitstats_only}. The target responses exhibit near-ideal entropy values ($\approx 0.99$), confirming a balanced and unbiased distribution of response bits. Most models preserve similar entropy characteristics; however, GBNN shows a significant entropy drop, indicating biased predictions and mode collapse. This behavior further reinforces that the model overfits the training data without learning a stable or meaningful mapping.

\begin{table}[htbp]
\caption{Bit-level statistical analysis of predicted responses.}
\label{tab:bitstats_only}
\centering
\renewcommand{\arraystretch}{1.2}
\setlength{\tabcolsep}{4pt}
\footnotesize
\begin{tabular}{|l|c|c|c|c|c|}
\hline
\textbf{Entity} & \textbf{Mean} & \textbf{Var} & \textbf{Avg Entropy} & \textbf{Min Ent.} & \textbf{Max Ent.} \\
\hline
Target Y & 0.5501 & 0.2474 & 0.9926 & 0.9878 & 0.9983 \\
ANN      & 0.5584 & 0.2465 & 0.9898 & 0.9838 & 0.9960 \\
GBNN     & 0.5443 & 0.1516 & 0.6701 & 0.0544 & 0.9999 \\
XGBoost  & 0.5571 & 0.2467 & 0.9903 & 0.9834 & 0.9945 \\
DT       & 0.5466 & 0.2477 & 0.9935 & 0.9860 & 0.9969 \\
RF       & 0.5819 & 0.2431 & 0.9800 & 0.9600 & 0.9900 \\
\hline
\end{tabular}
\end{table}

To enable a fair comparison across heterogeneous ML and DL models, each progressing along a different training axis (epochs for ANN and GBNN, boosting rounds for XGBoost, tree depth for DT, and number of trees for RF, each model's original training axis is converted into a common \emph{normalised training progress} scale, $\tau \in [0,100]$.

Let $i \in \{1,2,\dots,N\}$ denote the original training-step index, where $N$ is the total number of recorded steps for a given model. The normalised step $\tau_i$ corresponding to step $i$ is defined as
\[
\tau_i = \frac{i-1}{N-1}\times 100, \qquad \tau_i \in [0,100].
\]
By this definition, the first step ($i=1$) maps to $\tau=0\%$, while the final step ($i=N$) maps to $\tau=100\%$, irrespective of the model's original training axis or the total number of steps.

Since accuracy values are recorded only at the discrete positions $\{\tau_1,\tau_2,\dots,\tau_N\}$, they are projected onto a uniform set of 101 query points, $s_k = k$ for $k=0,1,\dots,100$, using piecewise linear interpolation. For any query point $s_k$ satisfying $\tau_i \leq s_k < \tau_{i+1}$, the interpolated accuracy $\hat{A}(s_k)$ is computed as
\[
\hat{A}(s_k) = A_i + \frac{s_k-\tau_i}{\tau_{i+1}-\tau_i}\left(A_{i+1}-A_i\right),
\]
where $A_i$ and $A_{i+1}$ are the recorded accuracy values (training or validation) at the adjacent normalised steps $\tau_i$ and $\tau_{i+1}$ that bracket $s_k$. This procedure preserves the overall shape of each learning curve while producing directly comparable trajectories over a shared normalised domain for all five models.

\begin{figure}[htbp]
\centering
\includegraphics[width=\linewidth]{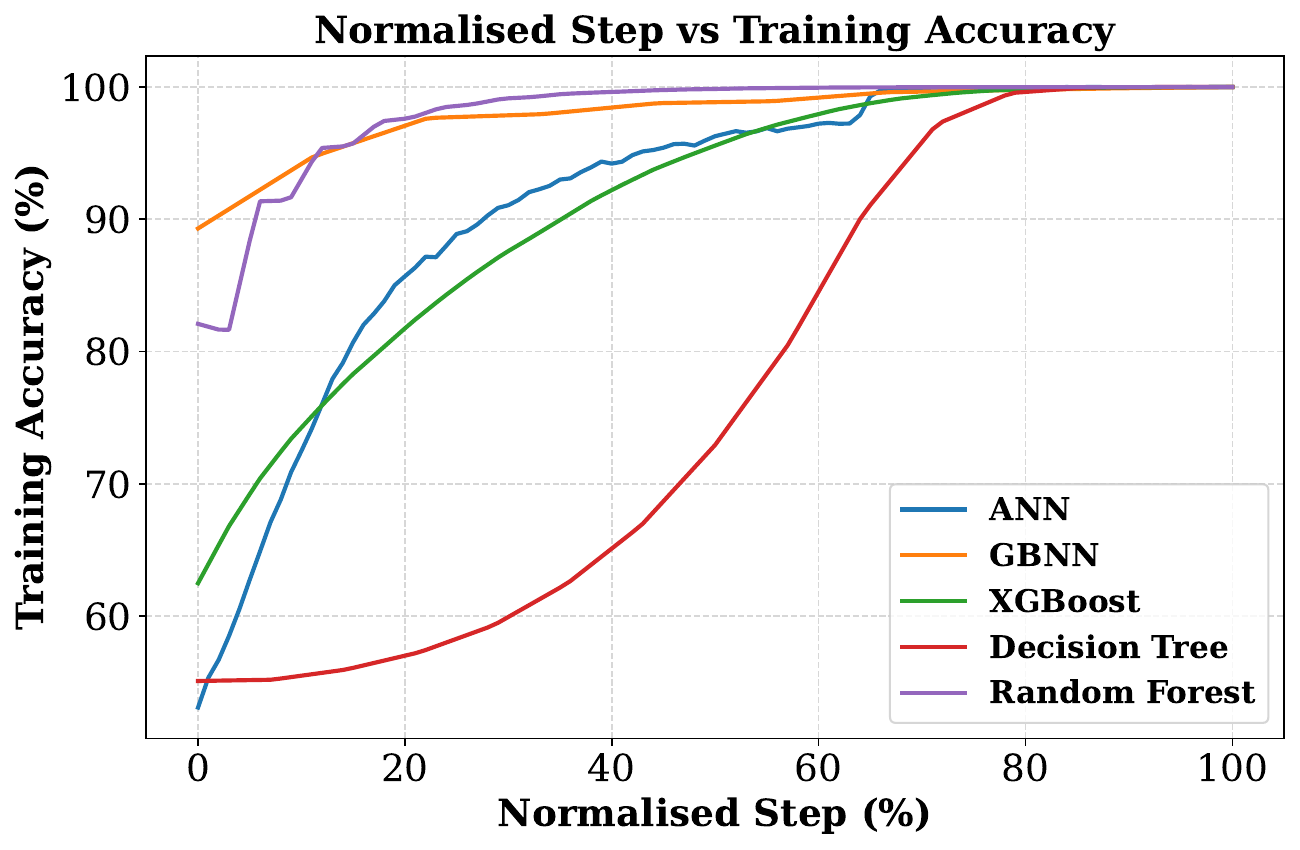}
\caption{Normalised step vs.\ training accuracy for all ML/DL models, showing convergence toward near-perfect training accuracy across ANN, GBNN, XGBoost, Decision Tree, and Random Forest.}
\label{fig:epoch-accuracy-train}
\end{figure}

\begin{figure}[htbp]
\centering
\includegraphics[width=\linewidth]{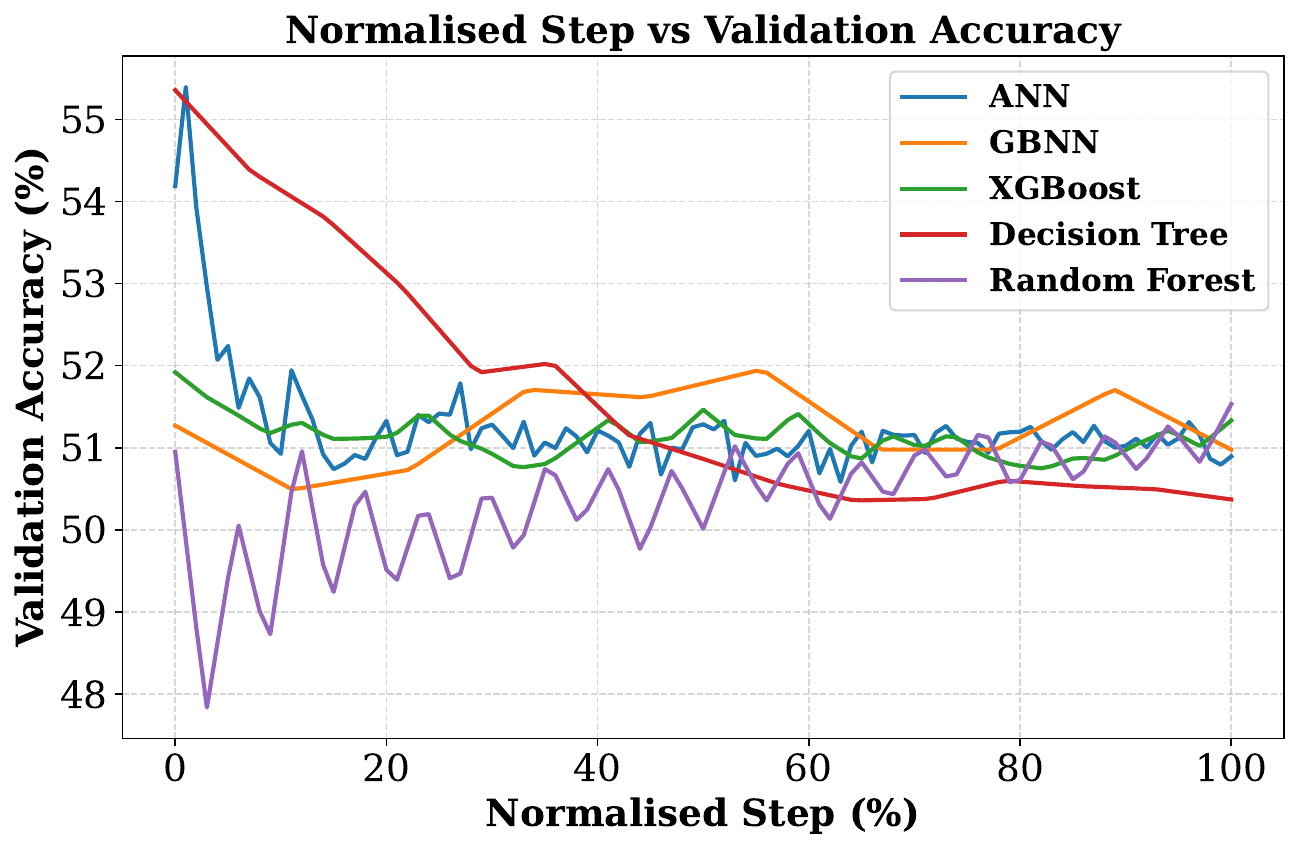}
\caption{Normalised step vs.\ validation accuracy for all ML/DL models, showing fluctuations close to the random-guess baseline across ANN, GBNN, XGBoost, Decision Tree, and Random Forest.}
\label{fig:epoch-accuracy-val}
\end{figure}

Figures~\ref{fig:epoch-accuracy-train} and~\ref{fig:epoch-accuracy-val} present the resulting training and validation trajectories on this common normalised scale. Figure~\ref{fig:epoch-accuracy-train} shows the normalised step versus training accuracy for all five evaluated models. As the normalised step increases, all models exhibit a strong upward trend and eventually converge to near-perfect training accuracy. This behaviour indicates that each learning algorithm has sufficient representational capacity to fit, and effectively memorise, the training CRPs.

Figure~\ref{fig:epoch-accuracy-val} presents the corresponding normalised step versus validation accuracy curves. In contrast to the training performance, the validation accuracy of all models remains confined to a narrow range around the random-guess baseline, fluctuating approximately near $50\%$ throughout the training process. Although a few models show higher or lower values during the early stage of training, these variations are not sustained, and no model demonstrates a meaningful improvement in generalisation. The persistent gap between near-perfect training accuracy and near-random validation accuracy provides strong evidence of overfitting and indicates that the learned behaviour is dominated by memorisation rather than extraction of a robust RC-PUF mapping.

\begin{figure}[htbp]
\centering
\includegraphics[width=\linewidth]{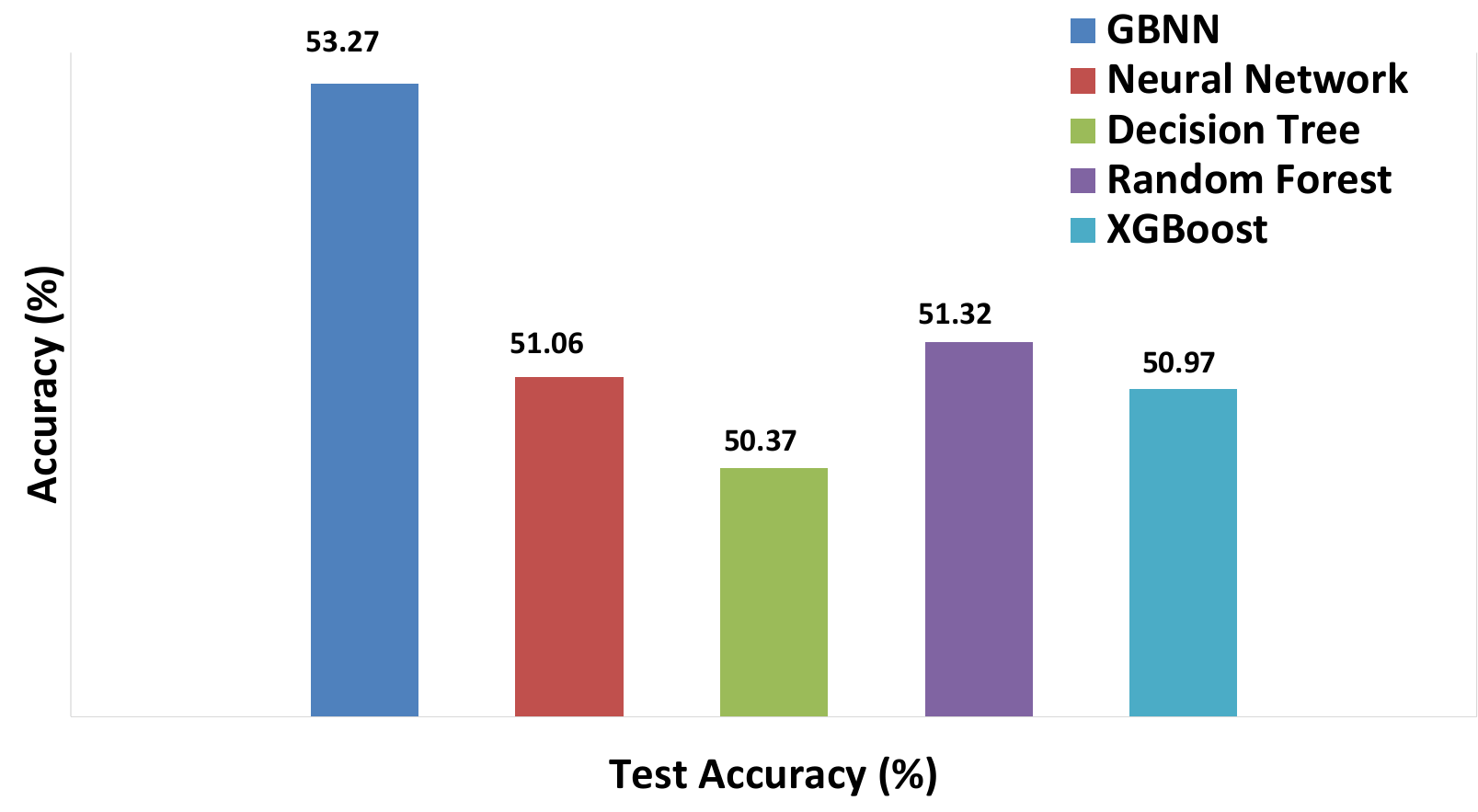}
\caption{Comparison of test accuracy across different ML/DL models for RC-PUF modeling attacks.}
\label{fig:test-accuracy-models}
\end{figure}

For a consolidated comparison, Figure~\ref{fig:test-accuracy-models} summarises the final test accuracy of all evaluated models using a bar chart. The test accuracy values remain concentrated within a narrow interval of approximately $50$--$53\%$, confirming that none of the considered ML or DL models is able to successfully model the RC-PUF response behaviour. This observation is consistent with the numerical results reported in Table~\ref{tab:perf_only} and further supports the conclusion that the proposed RC-PUF architecture exhibits strong resistance against modelling attacks.

\begin{table}[t]
\centering
\caption{Comparison of ML/DL Attack Resistance with Existing PUFs}
\label{tab:ml_comparison}
\renewcommand{\arraystretch}{1.2}
\setlength{\tabcolsep}{2.2pt}
\begin{tabular}{|l|l|c|c|c|}
\hline
\textbf{PUF} & \textbf{Model} & \textbf{Train (\%)} & \textbf{Test (\%)} & \textbf{Resistance} \\
\hline
Arbiter PUF \cite{10.1145/1866307.1866335} & LR/SVM & $\sim$100 & $>$90 & Low \\
\hline
XOR Arbiter PUF \cite{10.1145/1866307.1866335} & ANN & $\sim$100 & 70--85 & Moderate \\
\hline
RO PUF \cite{suh2007physical} & RF/ANN & $\sim$100 & 60--75 & Moderate \\
\hline
FF Arbiter PUF \cite{lim2005extracting} & SVM & $\sim$100 & 65--80 & Moderate \\
\hline
Proposed RC-PUF & Multi-ML/DL & \textbf{100} & \textbf{50--53} & \textbf{High} \\
\hline
\end{tabular}
\end{table}

Table~\ref{tab:ml_comparison} presents a comparative evaluation of the proposed RC-PUF against existing PUF architectures in terms of resistance to ML/DL-based modeling attacks. Conventional PUF designs, including Arbiter, XOR Arbiter, Ring Oscillator, and Feed-Forward Arbiter PUFs, exhibit relatively high test accuracy under machine learning models, indicating partial learnability of their challenge–response behavior. In contrast, the proposed RC-PUF demonstrates a clear disparity between training and testing performance, where all evaluated models achieve nearly 100\% training accuracy while the test accuracy remains confined to 50–53\% with zero exact-match rate. This behavior closely corresponds to random guessing and indicates the absence of a learnable mapping between challenges and responses. Therefore, the proposed dynamically reconfigurable RC-PUF exhibits strong resistance to modeling attacks and provides a more secure and reliable solution for IoT-based hardware authentication.

Overall, the combined numerical and visual results demonstrate that the proposed RC-PUF exhibits a high-entropy, non-stationary, and non-learnable challenge--response relationship. The inability of all tested ML/DL models to generalize beyond random guessing provides strong empirical evidence of robustness against modeling attacks. These findings validate that the RC-PUF's dynamic reconfiguration and analog delay characteristics ensure unpredictability, making it well-suited for secure hardware-based authentication in resource-constrained IoT environments.

\section{Conclusion}
This work assessed the resistance of an RC-based PUF design against multiple ML/DL modeling attacks including Decision Tree, Random Forest, XGBoost, ANN, and GBNN. All models achieved perfect training accuracy, confirming proper network convergence and training configuration. However, validation accuracy saturated near 50\%, with zero exact-match success across all architectures. This apparent ``failure'' of the models is, in fact, a strong indicator of PUF robustness. Entropy analysis revealed that both target and predicted responses exhibit near-unity entropy ($\approx 1$), implying that each bit is statistically independent and unpredictable. Consequently, the RC-PUF behaves as a high-entropy, dynamically reconfigurable system whose challenge--response mapping is non-deterministic and unlearnable. The results confirm that the proposed RC-PUF architecture provides inherent resistance to ML/DL-based modeling attacks and ensures a high degree of security for IoT hardware authentication applications.

\bibliographystyle{IEEEtran}
\bibliography{Reference}

\end{document}